\documentstyle[12pt]{article}
\begin{document}
\title{"THE UNIVERSE OF FLUCTUATIONS"}
\author{B.G. Sidharth\\ Centre for Applicable Mathematics \& Computer Sciences\\
B.M. Birla Science Centre, Hyderabad 500 063 (India)\\
Dedicated to the memory of my parents}
\date{}
\maketitle
\footnotetext{International Journal of Modern Physics 'A'}
\begin{abstract}
We discuss a recent model of a Quantum Mechanical Black Hole (QMBH) which
describes the most fundamental known particles the leptons and approximately
the quarks in terms of the Kerr-Newman Black Hole with a naked singularity
shielded by Zitterbewegung effects. This goes beyond the Zitterbewegung and
self interaction models of Barut and Bracken, Hestenes, Chacko and others
and provides a unified picture which amongst other things gives a rationale
for and an insight into:\\
1. The apparently inexplicable reason why complex space time transformations
lead to the Kerr-Newman metric in General Relativity.\\
2. The value of the fine structure constant.\\
3. The ratio between electromagnetic and gravitational interaction strengths.\\
4. The anomalous gyromagnetic ratio for the electron.\\
5. Why the neutrino is left handed.\\
6. Why the charge is discrete.\\
In the spirit of Effective Field Theories, this model provides an alternative
formalism for Quantum Theory and also for its combination with General
Relativity.\\
Finally a mechanism for the formation of these QMBH or particles is explored
within the framework of Stochastic Electrodynamics, QED and Quantum Statistical
Mechanics. The cosmological implications are then examined. It turns out that
a surprisingly large number of facts, including some which were hitherto inexplicable,
follow as a consequence of the model. These include a theoretical deduction of the
Mass, Radius and Age of the Universe, as also the values of Hubble's constant
and the Cosmological constant.
\end{abstract}
\section{Introduction}
In a previous communication$^1$ it was suggested that a typical elementary
particle, the electron can be considered to be what was termed a Quantum
Mechanical Black Hole (or QMBH), made up of a relativistic fluid of subconstituents,
described by the Kerr-Newman metric giving both its gravitational and
electromagnetic fields$^2$. It was pointed out that alternatively the QMBH could
be described as a relativistic vortex in the hydrodynamical formulation. It
was pointed out that the QMBH or vortex could also be thought of as a
relativisitc rotating shell.\\
In Section 2 we examine this model which explains several observed facts,
while in Section 3 we try to explore the mechanism which triggers off the
formation of these QMBH particles. In Section 4 we examine the cosmological
implications of the model and again discover that a surprisingly large
number of observed facts are neatly explained. Finally in Section 5 we
make some comments and observations.
\section{Quantum Mechanical Black Holes}
If we adhoc treat an electron as a charged and spinning black hole, described
by the Kerr-Newman metric, the pleasing fact which emerges is that this
metric describes the gravitational and electromagnetic field of an electron
including the anomalous gyromagnetic ratio$^2$, $g=2.$\\
However the horizon of the Kerr-Newman Black Hole becomes in this case
complex$^3$,
\begin{equation}
r_+ = \frac{GM}{c^2} + \imath b,b \equiv (\frac{G^2Q^2}{c^8} + a^2 -
\frac{G^2M^2}{c^4})^{1/2}\label{e1}
\end{equation}
where $G$ is the gravitational constant, $M$ the mass and $a \equiv L/Mc,L$
being the angular momentum. That is, we have a naked singularity apparently
contradicting the cosmic censorship conjecture. However, in the Quantum
Mechanical domain, (\ref{e1}) can be seen to be meaningful.\\
Infact, the position coordinate for a Dirac particle is
given by$^4$
\begin{equation}
x_\imath = (c^2p_\imath H^{-1} t + a_\imath)+\frac{\imath}{2}
c\hbar (\alpha_\imath - cp_\imath H^{-1})H^{-1},
\label{e2}
\end{equation}
where $a_\imath$ is an arbitrary constant and $c\alpha_\imath$ is the velocity operator
with eigen values $\pm c$. The real part in (\ref{e2}) is the usual position
while the imaginary part arises from Zitterbewegung. Interestingly, in both
(\ref{e1}) and (\ref{e2}), the imaginary part is of the order of $\frac{\hbar}
{mc}$, the Compton wavelength, and leads to an immediate identification of
these two equations. We must remember that our physical measurements are
gross - they are really measurements averaged over a width
of the order $\frac{\hbar}{mc}$. Similarly, time measurements are imprecise
to the tune $\sim \frac{\hbar}{mc^2}$. Very precise measurements if possible,
would imply that all Dirac particles would have the velocity of light, or
in the Quantum Field Theory atleast of Fermions, would lead to divergences.
(This is closely related to the non-Hermiticity of position operators in
relativistic theory as can be seen from equation (\ref{e2}) itself$^5$.
Physics begins after an averaging over the above unphysical
space-time intervals. In the process as is known (cf.ref.5), the imaginary
or non-Hermitian part of the position operator in (\ref{e2}) disappears.
That is in the case of the QMBH (Quantum Mechanical Black Hole), obtained by
identifying (\ref{e1}) and (\ref{e2}), the naked singularity is shielded
by a Quantum Mechanical censor.\\
To examine this situation more closely we reverse the arguments after
equation (\ref{e2}), and consider instead the complex displacement,
\begin{equation}
x^\mu \to x^\mu + \imath a^\mu\label{e3}
\end{equation}
where $a^o \approx \frac{\hbar}{2mc^2}, \mbox{and} a^\mu \approx \frac{\hbar}{mc}$
as before. That is, we probe into the QMBH or the Zitterbewegung region
inside the Compton wavelength as suggested by (\ref{e1}) and (\ref{e2}).
Remembering that $|a^\mu| < < 1$, we have, for the wave function,
\begin{equation}
\psi (x^\mu) \to \psi (x^\mu + \imath a^\mu) = \frac{a^\mu}{\hbar}
[\imath \hbar \frac{\partial}{\partial x^\mu} + \frac{\hbar}{a^\mu}]
\psi (x^\mu)\label{e4}
\end{equation}
We can identify from (\ref{e4}), by comparison with the well known electromagnetism-
momentum coupling, the usual electrostatic charge as,
\begin{equation}
\Phi e = \frac{\hbar}{a^o} = mc^2\label{e5}
\end{equation}
In the case of the electron, we can verify that the equality (\ref{e5})
is satisfied. Infact it was shown that from here we can get a rationale
for the value of the fine structure constant (cf.ref.1).\\
We next consider the spatial part of (\ref{e3}), viz.,
$$\vec x \to \vec x + \imath \vec a, \mbox{where} |\vec a| = \frac{\hbar}{2mc},$$
given the fact that the particle is now seen to have the charge $e$ (and mass
$m$). As is well known this leads in General Relativity from the static Kerr
metric to the Kerr-Newman metric where the gravitational and electromagnetic
field of the particle is given correctly, including the anomalous factor
$g = 2$. In General Relativity, the complex transformation (\ref{e3}) and
the subsequent emergence of the Kerr-Newman metric has no clear explanation. Nor
the fact that, as noted by Newman$^6$ spin is the orbital angular momentum
with an imaginary shift of origin. But in the above context we can see the
rationale: the origin of (\ref{e3}) lies in the QMBH and Zitterbewegung
processes inside the Compton wavelength.\\
However the following question has to be clarified: How can an electron
described by the Quantum Dirac spinor $(\theta_\chi)$, where $\theta$
denotes the positive energy two spinor and $\chi$ the negative energy
two spinor, be identified with the geometrodynamic Kerr-Newman Black
Hole characterised by the curved space time (without any doublevaluedness,
cf.ref.2).\\
We observe that as is well known,$^7$ at and within the Compton wavelength
it is the negative energy $\chi$ that dominates. Further, under reflection,
while $\theta \to \theta, \chi$ behaves like a psuedo-spinor,\\
$$\chi \to -\chi$$
Hence the operator $\frac{\partial}{\partial x^\mu}$ acting on $\chi$, a
density of weight $N = 1,$ has the following behaviour$^8$,
\begin{equation}
\frac{\partial \chi}{\partial x^\mu} \to \frac{1}{\hbar} [\hbar \frac{\partial}
{\partial x^\mu} - NA^\mu]\chi\label{e6}
\end{equation}
where,
\begin{equation}
A^\mu = \hbar \Gamma_\sigma^{\mu \sigma} = \hbar \frac{\partial}{\partial x^\mu}
log (\sqrt{|g|})\label{e7}
\end{equation}
As before we can identify $NA^\mu$ in (\ref{e6}) with the electromagnetic
four potential. That $N = 1$, explains the fact that charge is discrete.\\
In this formulation, electromagnetism arises from the covariant derivative
which is the result of the Quantum Mechanical behaviour of the negative
energy components of the Dirac spinor at the Compton wavelength scale. We can see
at once how an electron can be associated with curvature and how the double
connectivity of spin half surfaces in the geometrodynamical formulation.
(\ref{e7}) strongly resembles Weyl's formulation for the unification
of electromagnetism and gravitation$^9$. However it must be noted that the
original Christofell symbol of Weyl contained two independent entities viz.
the metric tensor \underline{and} the electromagnetic potential, so that there was
really no unification. In our formulation we have used only the Quantum
Mechanical psuedo spinor property.\\
So we could treat the Quantum Mechanical Black Hole as a relativistic fluid
of subconstituents (or Ganeshas). In a linearized theory (cf.ref.2) we
have
\begin{equation}
g_{\mu v} = \eta_{\mu v} + h_{\mu v}, h_{\mu v} = \int \frac{4T_{\mu v}(t -
|\vec x - \vec x'|, \vec x')}{|\vec x - \vec x'|}d^3 x'\label{e8}
\end{equation}
It was then shown, (cf.ref.1), that not only do we recover the Quantum
Mechanical spin but using equation (\ref{e7}) and (\ref{e8}) that for
$r = |\vec x| >> |\vec x'|$ we get
\begin{equation}
\frac{e'e}{r} = A_o \sim \frac{\hbar c^3}{r} \int \rho \omega d^3 x'
\sim (Gmc^3)\frac{mc^2}{r}\label{e9}
\end{equation}
where $e' = 1 \mbox{esu}$ corresponds to the charge $N = 1$ and $e$ is the
test charge. (\ref{e9}) is correct and infact leads to the well known empirical result,
\begin{equation}
\frac{e^2}{Gm^2} \sim 10^{40},\label{e10}
\end{equation}
The above model gives a rationale for the left handedness of the neutrino,
which can be treated as an electron with vanishing mass so that the
Compton wavelength becomes arbitrarily large. For such a particle, we
encounter in effect the region within the Compton wavelength with the pseudo
spinorial property discussed above, that is left handedness.\\
Finally it may be remarked that the electron, the positron and its special case the neutrino
are the fundamental elementary particles which could be used to generate
the mass spectrum of elementary particles$^{10}$.\\
We now briefly examine why the Compton wavelength emerges as a fundamental
length. Our starting point could be the Dirac or Klein-Gordon equations. For
simplicity we consider the Klein-Gordon equation. It is well known that the
position operator is given by$^{5}$
\begin{equation}
\vec X_{op} = \vec x_{op} - \frac{\imath \hbar c^2}{2} \frac{\vec p}{E^2}\label{e11}
\end{equation}
(The Dirac equation also has a similar case).\\
We saw in (ref.1) that the imaginary part in equation (\ref{e11}) which
makes $\vec X_{op}$ non-Hermitian, and for the Dirac particle gives Zitterbewegung
disappears on averaging over intervals $\Delta t \sim \frac{\hbar}{mc^2} (\mbox{and}
\Delta r \sim \frac{\hbar}{mc})$ so that $\vec X_{op}$ becomes Hermitian (this is
also the content of the Foldy-Wothuysen transformation). Our physics, as
pointed out begins after such an average or Hermitization. Our measurement
in other words are necessarily gross to this extent - we will see this more
clearly. From equation (\ref{e11})
we now get
\begin{equation}
\hat X^2_{op} \equiv \frac{2m^3c^4}{\hbar^2} X^2_{op} = \frac{2m^3c^6}{\hbar^2}
x^2 + \frac{p^2}{2m}\label{e12}
\end{equation}
Mathematically equation (\ref{e12}) shows that $\hat X^2_{op}$ gives a problem
identical to the harmonic oscillator with quantized levels: Infact the quantized
"space-levels" for $\vec X^2_{op}$ turn out to be multiples of $(\hbar/mc)^2$!
From here, we get $\Delta t = \frac{\Delta x}{c} = \frac{\hbar}{mc^2}$.
\section{The formation of QMBH particles}
We now investigate how such QMBH can be formed. For this we digress
temporarily to vaccuum fluctuations. It is well known that there is a zero
point field (ZPF). According to QFT this arises due to the virtual quantum
effects of the electromagnetic field already present. Whereas according to
what has now come to be called Stochastic Electrodynamics (SED), it is
these ZPF that are primary and result in Quantum Mechanical primary effects
$^{11}$. Many Quantum Mechanical effects can indeed be explained this way.
Without entering into the debate about the ZPF fluctuations for the moment, we observe that
the energy of the fluctuations of the magnetic field in a region of length
$\lambda$ is given by$^{2}$ $(\vec E$ and $\vec B$ are electromagnetic field
strengths)
\begin{equation}
B^2 \sim \frac{\hbar c}{\lambda^4}\label{e13}
\end{equation}
If $\lambda$ as in the QMBH is taken to be the Compton wavelength,
$\frac{\hbar}{mc}$ (\ref{e13}) gives us for the energy in this volume
of the order $\lambda^3$,\\
$$\mbox{Total\quad energy\quad of\quad QMBH\quad}\sim \frac{\hbar c}{\lambda} =
mc^2,$$
exactly as required. In other words the entire energy of the QMBH
of mass $m$ can be thought to have been generated by the fluctuations
alone. Further the fluctuation in curvature over the length $l$ is given
by$^{2}$,
\begin{equation}
\Delta R \sim \frac{L^*}{l^3},\label{e14}
\end{equation}
where $L^*$ is the Planck length of the order $10^{-33}cms$.\\
For the electron which we consider, $l$ is of the order of the Compton
wavelength, that is $10^{-11}cms$. Substitution in (\ref{e14}) therefore
gives
$$\Delta R \sim 1$$
In other words the entire curvature of the QMBH is also generated by these
fluctuations. That is the QMBH can be thought to have been created by
these fluctuations alone.\\
Within the framework of QED, we can come to this conclusion in another
way$^{12}$. It is known that the vaccuum energy of the electron field with
a cut off $k_{max}$ is given by,
\begin{equation}
\frac{\mbox{Energy}}{\mbox{Volume}} \sim \hbar c k^4_{max}\label{e15}
\end{equation}
This is the same as equation (\ref{e13}) encountered earlier. Also the
infinite energy of the vaccuum is avoided by the assumption of the cut
off normally taken to be of the order of a typical Compton wavelength on
the ground that we do not know that the laws of electromagnetism are
valid beyond these high frequencies, that is within these length scales.\\
But the preceding discussion shows that it is natural to take $k_{max} =
\frac{mc}{\hbar},$ the inverse Compton wavelength of the electron. The energy
of the electron from equation (\ref{e15}) then comes out to be
$$E \sim mc^2,$$
as before. So we are led to the important conclusion that the
infinity of QED is avoided by the fact that QMBH are formed, rather than by
the arbitrary prescription of a cut off. Infact there is a further bonus
and justification for the above interpretation. Let us use in (\ref{e15})
the pion Compton wavelength as the cut off. The reason we choose the pion
is that it is considered to be a typical elementary particle in the sense
that it plays a role in the strong interactions, and further it could
be used as a building block for developing a mass spectrum, and finally
as seen in (ref.1) can be considered to be made up of an electron and
a positron. Then from (\ref{e15}) we can recover the pion mass,
$m_\pi$ and moreover,
\begin{equation}
Nm_\pi = M,\label{e16}
\end{equation}
where $N$ is the number of elementary
particles, typically pions, $N \sim 10^{80}$ and $M$ is the mass
of the universe,viz. $10^{56}gms$.\\
In other words, in our interpretation we have not only avoided the QED
infinity but have actually recovered the mass of the universe. We will return
to this point shortly.\\
We now consider the same scenario from a third point of view, viz. from the
standpoint of Quantum Statistical Mechanics. Here also the spirit is that
of randomness$^{13}$. A state can be written as
\begin{equation}
\psi = \sum_{n} c_n \phi_n,\label{e17}
\end{equation}
in terms of basic states $\phi_n$ which
could be eigen states of energy for example, with eigen values $E_n$. It
is known that (\ref{e17}) can be written as
\begin{equation}
\psi = \sum_{n} b_n \phi_n\label{e18}
\end{equation}
where $|b_n|^2 = 1$ if $E<E_n<E+\Delta$, and $= 0$ otherwise under the assumption
\begin{equation}
\overline{(c_n,c_m)} = 0, n \ne m\label{e19}
\end{equation}
(Infact $n$ could stand for not a single state but for a set of states
$n_\imath$ and so also $m$). Here the bar denotes a time average over a
suitable interval. This is the well known Random Phase Axiom and arises
due to the total randomness amongst the phases $c_n$. Also the expectation
value of any operator $O$ is given by
\begin{equation}
<O> = \sum_{n} |b_n|^2 (\phi_n, O \phi_n)/ \sum_{n} |b_n|^2\label{e20}
\end{equation}
(\ref{e18}) and (\ref{e20}) show that effectively we have incoherent states
$\phi_1, \phi_2,....$ once averages over time intervals for the phases
$c_n$ in (\ref{e19})vanish owing to their relative randomness.\\
In the light of the preceding discussion of random fluctuations in the context of QMBH
in SED and QED, we can interpret the above meaningfully: We can identify
$\phi_n$ with the ZPF. The time averages are the Zitterbewegung averages
over intervals $\sim \frac{\hbar}{mc^2}$. We then get disconnected or
incoherent particles or QMBH from a single background of vaccuum
fluctuations exactly as before. The incoherence arises because of the well
known random phase relation (\ref{e19}) that is after averaging over the
suitable interval.\\
But in all of the above considerations, and in present day theory the question
that comes up is: How can we reconcile the fact that the various particles
in the universe are not infact incoherent but rather occupy a single
coherent space-time. The answer which can now be seen to emerge in the
light of the above discussion is that all these particles are linked by
interaction. These interactions as pointed out in (ref.1) arise within
the Compton wavelength or Zitterbewegung region, that is in phenomena
within the time scale $\frac{\hbar}{mc^2}$. It will be observed in the
above discussion that at these time scales the equation (\ref{e19}) is
no longer valid and we have to contend with equation (\ref{e17}) rather
than equation (\ref{e18}). So interactions arising within the
Compton wavelength link or make coherent
the various particles.\\
Infact all this is perfectly in tune with the QFT picture wherein the
interactions are caused by virtual particles with life time less than
$\frac{\hbar}{mc^2}$. It may also be observed that in Wheeler's Geometrodynamical
model$^{14}$, the various particles are linked by exactly such
wormholes linking distant regions.\\
In the above formulation we could take $\phi_n$ to be the particlets or Ganeshas instead of
energy eigen states, that is to be position eigen states and consider sets
$$\overline{(c_{n_\imath}, c_{m_j})} = 0,$$
exactly as before (cf.(\ref{e19})). Each set $\phi_{n_\imath}$ defines a particle
$P_n$ consisting of $n_\imath$ Ganeshas or particlets. It is the link
at $\Delta t < \hbar/mc^2$ between $P_n$ and $P_m$ which puts otherwise incoherent particles into a
single space-time, that is allows interactions.\\
In other words, a set of particles can be said to be in the same space-time
if every particle interacts with atleast one other member of the set.\\
For completeness we mention that the above bunching could be carried out
in principle for two
or more universes. Thus a set of particles constitutes universe $U_1$ while
another set of particles constitute an incoherent universe $U_2$. Again the
incoherence can be broken at a suitable time scale (cf.ref.15 for a pictorial
model in terms of wormholes).\\
There is another way of looking at all this.
We first note that the space-time symmetry of relativity has acquired
a larger than life image. Infact our perception of the universe is
essentially one of all space (or as much of it as possible) at one instant
of time (cf.also ref.2). Further, time is essentially an ordering or sequencing
of events. To understand time we must know on what basis this ordering is
done so that causality and other laws of physics hold or in other words we
have the universe of the physical hyperboloid.\\
We now approach this problem by trying to liberate the sequence of
events in time from any ordering at all. At first sight it would appear that
this approach would lead to a chaotic universe without physics that is
causality, interaction and so on. We will actually try to attempt to explain
the emergence of physics from such a, what may be called pre-space-time scenario. It must be
noted that both Special and General Relativity work in a deterministic
space-time. Even relativistic Quantum Mechanics and Quantum Field Theory
assume the space-time of Special Relativity. Quantum Gravity on the other
hand which has not yet proved to be a completely successful theory questions
this concept of space-time$^{16}$.\\
While a random time sequence is ruled out at what may be called the macro
level, in our case above the Compton wavelength, within the framework of
QMBH and as seen above this is certainly possible below the Compton
wavelength scale. Infact this is the content of non locality and non Hermiticity
of the Zitterbewegung in the region of QMBH.\\
So we start with truly instantaneous point particles or particlets (or Ganeshas)
which are therefore indistinguishable, (cf. ref.1) (and could be denoted by
$\phi_n$ of (\ref{e17})). We then take a random sequence of such
particlets$^{13}$. Such a sequence for the interval $\Delta t \sim \frac
{\hbar}{mc^2}$ in time collectively constitutes a particle that has come
into existence and is spread over a space interval of the order of the
Compton wavelength. In other words we have made a transition from pre-
space-time to a particle in space-time. This is exactly the averaging over random phases in
equation (\ref{e19}). Hermiticity of position operators has now been
restored and we are back with the states $\phi_n$ in equation
(\ref{e18}). All this is in the spirit that our usual time is such that,
with respect to it vaccuum fluctuations are perfectly random as pointed
by Macrea$^{17}$. So the subconstituents of the relativistic fluid given
in (\ref{e8}) (or the Quantized Vortex in the hydrodynamical formulation
(cf. ref.1)), are precisely these particlets.\\
To visualize the above consideration in greater detail we first consider strictly
point particles obtained by taking the random sequences over time
intervals $\sim \frac{\hbar}{mc^2}$. We consider an assembly of such truly
point particles which as yet we cannot treat either with Fermi-Dirac or
Bose-Einstein statistics but rather as a Maxwell-Boltzman distribution. If there
are $N$ such particles in a volume $V$, it is known that$^{13}$, the volume
per particle is of the order of,
$$(\frac{V}{N})^{1/3} \sim \lambda_{thermal} \approx \frac{\hbar}{\sqrt{m^2c^2}} =
\frac{\hbar}{mc},$$
where we take the average velocity of each particle to be equal to $c$. Infact,
this is exactly what happens, as Dirac pointed out (cf. ref.4), for a truly
hypothetical point electron, in the form of Zitterbewegung within the
Compton wavelength.\\
So the Compton wavelength arises out of the (classical) statistical inability to
characterise a point particle precisely: It is not that the particle has an
extension per se. In this sense the Compton wavelength has a very
Copenhagen character, except that it has been deduced on the basis of an
assembly of particles rather than an isolated particle.
\section{The Universe of Fluctuations}
The question that arises is, what are the cosmological implications of the
above scenario, that is, if we treat the entire universe as arising from
fluctuations, is this picture consistent with the observed universe? It
turns out that not only is there no inconsistency, but on the contrary
a surprising number of correspondences emerge.\\
The first of these is what we have encountered a little earlier viz.
the fact that we recover the mass of the universe as in equation (\ref{e16}).\\
We can next deduce another correspondence. The ZPF gives the correct
spectral density viz.
$$\rho (\omega) \alpha \omega^3$$
and infact the Planck spectrum$^{18}$. We then get the total intensity
of radiation from the fluctuating field due to a single star as$^{11}$,
$$I(r) \alpha \frac{1}{r^2}$$
It then follows that given the observed isotropy and homogeneity of the
universe at large, as is well known,
\begin{equation}
M \alpha R\label{e21},
\end{equation}
where $R$ is the radius of the universe.\\
Equation (\ref{e21}) is quite correct and infact poses a puzzle, as is
well known and it is to resolve this dependence that dark matter has
been postulated$^{19}$ whereas in our formulation the correct mass radius
dependence has emerged quite naturally without any other adhoc postulates.\\
As we have seen above the Compton wavelength of a typical particle, the pion viz $l_\pi$ can be
given in terms of the volume of uncertainity. However in actual observation
there is an apparent paradox. If the universe is $n$ dimensional then we
should have,
$$Nl^n_\pi \sim R^n$$
for the universe itself. This relation is satisfied with $n = 2$ in which
case we get a relation that has been known emperically viz.,
$$l_\pi \sim \frac{R}{\sqrt{N}}$$
(Even Eddington had used this relation).\\
So in conjunction with (\ref{e16}) we have an apparent paradox where the actual universe appears to be
two dimensional. This will be resolved shortly and it will be seen that
there is no contradiction.\\
Another interesting consequence is as follows: According to our formulation
the gravitational potential energy of a pion in a three dimensional isotropic
sphere of pions is given by
$$\frac{Gm_\pi M}{R}$$
This should be equated with the energy of the pion viz. $m_\pi c^2$. We then
get,
\begin{equation}
\frac{GM}{c^2} = R,\label{e22}
\end{equation}
a well known and observationally correct relation. In our formulation we
get $m_\pi$ from the ZPF and given $N$ we know $M$ so that from equation
(\ref{e22}) we can deduce the correct radius $R$ of the universe.\\
Proceeding further we observe that the fluctuations in the particle number $N$ itself is
of the order $\sqrt{N}^{13,20}$. Also $\Delta t$ above is the typical
fluctuating time. So we get,
$$\frac{dN}{dt} = \frac{\sqrt{N}}{\Delta t} = \frac{m_\pi c^2}{\hbar} \sqrt{N}$$
whence as $t = 0, N = 0,$
\begin{equation}
\sqrt{N} = \frac{2m_\pi c^2}{\hbar} .T\label{e23}
\end{equation}
where $T$ is the age of the universe $\approx 10^{17}secs$. It is remarkable
that equation (\ref{e23}) is indeed correct. One way of looking at this is
that not only the radius but also the age of the universe is correctly
determined. As we saw before,
$$R = \frac{GM}{c^2} = \frac{GNm_\pi}{c^2}$$
so that
\begin{equation}
\frac{dR}{dt} = \frac{Gm_\pi}{c^2} \frac{dN}{dt} = \frac{Gm^2_\pi}{\hbar} \sqrt{N} = HR\label{e24}
\end{equation}
where
\begin{equation}
H = \frac{Gm_\pi^3 c}{\hbar^2}\label{e25}
\end{equation}
One can easily verify that (\ref{e25}) is satisfied for the Hubble constant so
that (\ref{e24}) infact gives the Hubble's velocity distance relation.\\
Furthermore from (\ref{e25}) we deduce that,
\begin{equation}
m_\pi = (\frac{\hbar^2 H}{Gc})^{1/3}\label{e26}
\end{equation}
It is remarkable that equation (\ref{e26}) is known to be true from a
purely empirical standpoint$^{21}$. However we have actually deduced it in
our formalism. Another way of interpreting equation (\ref{e26}) is that
given $m_\pi$ (and $\hbar, G$ and $c$) we can actually deduce the value of
$H$ in our formalism.\\
From Equation (\ref{e24}), we deduce that,
\begin{equation}
\frac{d^2R}{dt^2} = H^2R\label{e27}
\end{equation}
That is, effectively there is a cosmic repulsion. Infact, from (\ref{e27}) we
can identify the cosmological constant as
$$\Lambda \sim H^2$$
which is not only consistent but agrees exactly with the limit on this constant
(cf.ref.2).\\
The final correspondence is to do with an explanation for the microwave
cosmic background radiation within the above framework of fluctuations. It
is well known that the fluctuations of the Boltzmann $H$ function for interstellar
space is of the order of $10^{-11}secs^{13}$. These fluctuations can be
immediately related to the ZPF exactly as in the case of the Lamb shift
(cf.ref.2). So $\frac{\hbar}{mc^2} = 10^{-11}$ or the associated wavelength
viz.,
$$\frac{\hbar}{mc} \sim 0.3 cms,$$
which corresponds to the cosmic background
radiation$^{3}$. The same conclusion can be drawn from a statistical
treatment of interstellar Hydrogen$^{22}$.
\section{Comments}
1. We could arrive at equation (\ref{e13}) by a slightly different route
(cf.ref.2). We could start with a single oscillator in the ground state
described by the wave function
\begin{equation}
\psi(x) = \mbox{const} \quad exp[-(m\omega/2 \hbar)x^2]\label{e28}
\end{equation}
which would fluctuate with a space uncertainity of
$$\Delta x \sim (\hbar/m\omega)^{\frac{1}{2}} = \frac{\hbar}{mc}$$
The electromagnetic ZPF could be treated as an infinite collection of
independent oscillators and we could recover equation (\ref{e13}).\\
2.Earlier we skirted the issue whether the ZPF is primary or secondary.
We now start either with the ZPF or with the pre-space-time background
field of the instantaneous particles (or Ganeshas). We could assign a
probability $p$ for them to appear in space-time and the probability
$1-p = q$ for this not to happen. From here we get the probability for
$N$ of them to appear as
\begin{equation}
\mbox{Probability}\quad \alpha \quad exp \quad [-\mu^2 N^2]\label{e29}
\end{equation}
This immediately ties up with the considerations following from equation
(\ref{e12}) (cf.ref.2), if we identify $N$ with $x$. The justification
for this can be seen by a comparison with $|\psi (x)|^2$ from (\ref{e28}):
From (\ref{e29}), the probability is non-negligible if
$$\Delta N \sim \frac{1}{\mu},$$
which turns out to be, from (\ref{e28}),
$$\Delta N \sim \frac{1}{\mu} \approx \frac{\hbar}{mc},$$
the Compton wavelength. Thus once again we conclude from (\ref{e29})
that a probabilistic fluctuational collection of instantaneous particlets
from a pre-space-time background shows up as a particle in space-time.\\
Wheeler considers the algebra of propositions as providing the link
between what he terms pre-geometry and geometrodynamics. In our formulation
probabilistic fluctuations lead to space-time and physics from
pre-space-time.\\
3. It was pointed out that the equation
\begin{equation}
l_\pi \sim \frac{R}{\sqrt{N}}\label{e30}
\end{equation}
suggests that the universe is apparently two dimensional. This paradoxical
result is consistent with astrophysical data (cf.ref.19). We could resolve
the paradox as follows:\\
We start with the fact that the universe on the average is neutral. Further
the fluctuation in the number of electrons is $\sim \sqrt{N}$. So an extra
electrostatic potential energy is created which is balanced by (or in
our formulation manifests itself as) the energy of the electron itself
(cf.ref.20):
$$\frac{e^2\sqrt{N}}{R} = mc^2$$
which leads to the above relation.\\
So in the conventional theory, that is in the language of a fixed particle
number universe, we would say that the universe is apparently two dimensional.
But once we recognise the fluctuations, the universe is really three
dimensional. Infact the fundamental equation (\ref{e10}) which was
derived purely from the point of view of an isolated particle can also be
derived on the basis of a "two dimensional" universe$^{23}$.\\
4. The considerations of the previous section show that there exists, what
may be called a micro-macro nexus: Fundamental constants of Quantum Theory
are tied up with constants from macro physics and cosmology. So the universe
is holistic. It has a slightly different connotation from the Machian formulation,
because the latter deals with a deterministic universe with rigid physical
laws.\\
Infact from (\ref{e30}), (\ref{e22}) and (\ref{e23}), we can deduce that,
\begin{equation}
\frac{2Gm_\pi^3c}{\hbar^2} = \frac{1}{T}\label{e31}
\end{equation}
which is a variant of equation (\ref{e10}), if we replace its right side
by $\sqrt{N}$. This may be interpreted as giving $e,G,c$ or $\hbar$ in
terms of $m_\pi\quad \mbox{and}\quad N$. More interestingly, (\ref{e31})
gives the variation of $G$, or more generally, the left side, with the
age of the universe (cf.ref.2 for Dirac's conjecture in this connection).\\
5. The quantization formula for space following from equation (\ref{e12}),
reflects an empirical formula deduced by Chacko$^{24}$ which can be used to
generate a mass spectrum. It also vindicates a close connection between
energy and space-time: As pointed out (cf.ref.1) inertial mass arises from
the non local Zitterbewegung processes within the Compton wavelength$^{25}$.\\
6. Finally we observe that inspite of similarities, the above scenario of
fluctuations differs from steady-state cosmology and the $C$ field
formulation$^{26}$.

\end{document}